\documentclass[referee]{aa}
\usepackage{graphicx}
\usepackage{txfonts}
\usepackage{cases}

\begin{document}
   \title{Chromospheric evaporation in sympathetic coronal bright points}

   \author{Q. M. Zhang\inst{1,2} \and
           H. S. Ji\inst{1}
          }

   \institute{Key Laboratory for Dark Matter and Space Science, Purple
              Mountain Observatory, CAS, Nanjing 210008, China \\
              \email{zhangqm@pmo.ac.cn}
              \and
              Key Lab of Modern Astronomy and Astrophysics, Ministry of
              Education, China\\
              }

   \date{Received; accepted}
    \titlerunning{Chromospheric evaporation in coronal bright points}
    \authorrunning{Zhang \& Ji}

  \abstract
   {Chromospheric evaporation is a key process in solar 
   flares that has extensively been investigated using the spectroscopic
   observations. However, direct soft X-ray (SXR) imaging of the process
   is rare, especially in remote brightenings associated with the primary 
   flares that have recently attracted dramatic attention.}
   {We intend to find the evidence for chromospheric evaporation 
   and figure out the cause of the process in sympathetic coronal bright 
   points (CBPs), i.e., remote brightenings induced by the primary CBP.}
   {We utilise the high-cadence and high-resolution SXR 
   observations of CBPs from the X-ray Telescope 
   (XRT) aboard the Hinode spacecraft on 2009 August 23.}
   {We discover thermal conduction front propagating from the
   primary CBP, i.e., BP1, to one of the sympathetic CBPs, i.e., BP2 that is 
   60$\arcsec$ away from BP1. The apparent velocity of the thermal conduction  
   is $\sim$138 km s$^{-1}$. Afterwards, hot plasma flowed upwards into the 
   loop connecting BP1 and BP2 at a speed of $\sim$76 km s$^{-1}$, a clear 
   signature of chromospheric evaporation. Similar upflow was also observed 
   in the loop connecting BP1 and the other sympathetic CBP, i.e., 
   BP3 that is 80$\arcsec$ away from BP1, though 
   less significant than BP2. The apparent velocity of the upflow is $\sim$47
   km s$^{-1}$. The thermal conduction front propagating from BP1 to BP3 
   was not well identified except for the jet-like motion also originating from BP1.}
   {We propose that the gentle chromospheric evaporation in the 
   sympathetic CBPs were caused by thermal conduction originating from 
   the primary CBP.}


   \maketitle

\section{Introduction} \label{s-intro}

Solar flares are transient energy release via magnetic reconnection 
accompanied by localized plasma heating and particle accelerations 
(Svestka \cite{sves76}; Gan et al. \cite{gan91}; Ding \& Fang \cite{ding01}; 
Ji et al. \cite{ji06,ji07,ji08}; Guo et al. \cite{guo08}; 
Fletcher et al. \cite{fle11}; Ning \& Cao \cite{ning11}; 
Shibata \& Magara \cite{shi11}; Chen \cite{chen12}; 
Hao et al. \cite{hao12}). According to the standard flare model, i.e.,
CSHKP model (Carmichael \cite{car64}; Sturrock \cite{stu66}; 
Hirayama \cite{hir74}; Kopp \& Pneuman \cite{kop76}), substantial
nonthermal electrons (10--100 keV) propagate 
downwards along the reconnected magnetic field lines and collide with 
the dense chromosphere, creating impulsive hard X-ray (HXR) emissions 
and thermalizing the local plasma to 1--10 MK. Meanwhile, thermal
conduction propagates downwards from the super-hot reconnection region. 
The overpressure of the chromosphere propels hot plasma into the tenuous 
coronal loops that emit strong emissions in soft X-ray (SXR), a process 
entitled ``chromospheric evaporation'' (Neupert \cite{neu68}; 
Fisher et al. \cite{fis85a,fis85b,fis85c}; Emslie et al. \cite{ems92}; 
Allred et al. \cite{all05}). 
Fisher et al. (\cite{fis85b}) divided the chromospheric evaporation
into two types according to the inputted energy flux. If the energy flux
exceeds a critical value of $\sim$10$^{10}$ erg cm$^{-2}$ s$^{-1}$, then 
explosive evaporation takes place accompanied by blue-shifts at speed of 
hundreds of km s$^{-1}$ in the emission lines formed in the coronal temperature
and red-shifts at speed of tens of km s$^{-1}$ in the emission lines formed
in the transition region and upper chromosphere (Brosius \& Holman \cite{bro07}). 
Otherwise, gentle evaporation takes place accompanied by blue-shifts 
at speed of tens of km s$^{-1}$ in all emission lines (Brosius \& Phillips 
\cite{bro04}; Brosius \& Holman \cite{bro09}; Berkebile-Stoiser et al. 
\cite{ber09}). Both nonthermal electrons and thermal conduction have been 
reported to play an important role in the gentle evaporation 
(Milligan et al. \cite{mill06}; Milligan \cite{mill08}). 
Up to now, chromospheric evaporation have been observed by the space-borne 
spectroscopic imagers in extreme ultraviolet (EUV; Chen \& Ding \cite{chen10}; 
Li \& Ding \cite{li11}) and HXR (Liu et al. \cite{liu06}; 
Ning et al. \cite{ning09}) wavebands. In SXR, however, the process has 
seldom been reported due to the low spatial resolution, time cadence, 
and temperature sensitivity of the previous telescopes 
(Silva et al. \cite{sil97}). Thanks to the state-of-the-art X-ray
Telescope (XRT; Golub et al. \cite{golub07}) aboard Hinode 
(Kosugi et al. \cite{kos07}), it becomes possible to seek for direct 
evidence of chromospheric evaporation. Nitta et al. (\cite{nit12}) for 
the first time observed clear chromospheric evaporation upflows 
arising almost symmetrically from the footpoints of magnetic loops. 
Among the 13 small transient brightenings observed by Hinode/XRT, 
nearly 46\% had an average apparent velocity of 100 km s$^{-1}$, and 
the other 23\% were much faster. Nevertheless, magnetic reconnections 
that lead to the evaporation were not recognized in their limited
sample, and the cause of evaporation (thermal conduction or nonthermal
electrons) was not clarified. Milligan (\cite{mill08}) showed 
apparent flow of hot material along the {\it GOES} B-class flare loop.
The absence of detectable HXR emission coupled with lower upflow 
velocities suggests gentle chromospheric evaporation driven by
thermal conduction.

Due to the complexities and interconnections of magnetic fields in 
the solar atmosphere, flares may not occur independently. 
Occasionally, a flare in the primary region may induce another one 
in a remote region, which is called ``sympathetic flare'' 
(Richardson \cite{rich51}). The mechanism of energy transport 
between the two regions has extensively been investigated. It was 
proposed that the possible driving agents could be energetic particles, 
shock waves, and thermal conduction (Machado et al. 
\cite{mach88}). For example, Zhang et al. (\cite{zhang00}) studied 
both the magnetic topology and time delay between the initial and 
sympathetic flares, finding that heat conduction is responsible 
for the sympathetic flare. For flares associated with coronal mass 
ejections (CMEs), the erupting flux rope and its envelope magnetic 
field may reconnect with the overlying field lines, producing a bright 
ribbon (or sympathetic flare) in a remote site (Wang \cite{wang05}). 
Interestingly, Brosius \& Holman (\cite{bro07}) discovered explosive 
chromospheric evaporation in a remote solar flare-like transient.
Wang \& Liu (\cite{wang12}) found remote brightenings at a region 
far from the circular-ribbon flares. Deng et al. (\cite{deng13}) also
discovered a circular-ribbon flare with a remote brightening that is 
predicted in three-dimensional fan-spine reconnection.

Coronal bright points (CBPs) are long-lived (8 hr) small-scale 
(10$\arcsec$--40$\arcsec$) brightenings in the lower corona 
(Golub et al. \cite{golub77}; Priest et al. \cite{pri94}; 
Tian et al. \cite{tian08}).
Like big flares, they are also believed to be heated by magnetic 
reconnection. Sometimes, recurrent strong flashes appear in CBPs with 
significant brightness enhancement (Zhang et al. \cite{zqm12}).
If a CBP flash occurs at the footpoint of a large-scale magnetic loop,
chances are that another transient brightening, i.e., CBP, will be
induced at the remote site of the large-scale loop due to the energy 
flux transported by nonthermal electrons or thermal conduction along 
the loop, which is termed ``sympathetic CBP''. Such events
and the associated chromospheric evaporation, 
to our knowledge, have not been investigated yet. In this letter, 
we report our detection of chromospheric evaporation in sympathetic 
CBPs observed by Hinode/XRT. In Section~\ref{s-data}, we describe 
the data analysis and show the results. Discussion and conclusion are 
presented in Section~\ref{s-disc}.

\section{Data analysis and results} \label{s-data}

XRT has 9 filters to achieve a wide temperature coverage from 0.3 MK to 30 MK.
During 2009 August 22--23, it pointed to a high-latitude region close 
to the north polar coronal hole using the C\_poly filter that has 
maximum temperature response at $\log T\approx6.95$. The partial-frame 
(384$\arcsec\times384\arcsec$) observations lasted for more than 10 hr from 
15:00 UT on August 22 to 01:30 UT on August 23 with time cadence of 
32 s and spatial resolution of 2\farcs06.
During the observations, a bright point in the field-of-view 
experienced several strong flashes, the last of which occurred around 
01:00 UT on August 23 and triggered another two bright points.

\begin{figure}
\includegraphics[width=12cm,clip=]{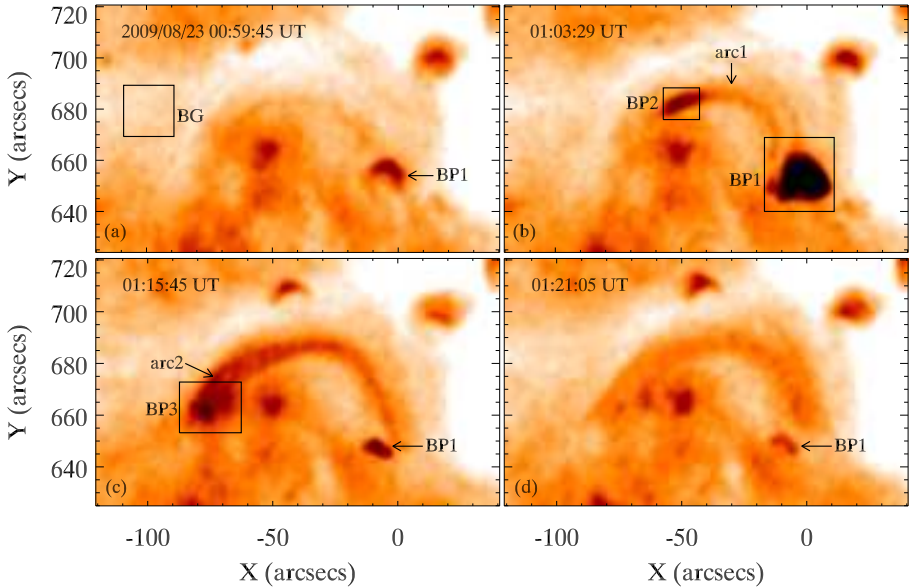}
\caption{Four snapshots of the SXR images observed by XRT C\_poly
filter. BG in panel {\bf a)} denotes the quiescent background region. 
BP1 represents the primary CBP. BP2 in panel {\bf b)} and BP3 in panel 
{\bf c)} stand for the sympathetic CBPs induced by BP1. Arc1 in 
panel {\bf b)} is the loop connecting BP1 and BP2. Arc2 in panel 
{\bf c)} takes root in BP3. Note that the color is reversed.}
\label{fig1}
\end{figure}

Figure~\ref{fig1} demonstrates four snapshots of the SXR images that
represent the four stages of the evolution of the CBPs, respectively.
Note that the color is reversed so that the dark regions in the panels
stand for bright features in the corona. In Fig.~\ref{fig1}a, the main 
bright point, i.e., BP1, exists as a tiny bright loop. The distance of 
the footpoints of the loop is about 15$\arcsec$. It started 
to increase slowly in size and brightness from $\sim$00:56 UT on
August 23. The quiescent square region, i.e., BG, within the box is the 
background with slight intensity fluctuation. The size and intensity
of BP1 increased rapidly from $\sim$01:00 UT and reached maximum at
$\sim$01:06 UT, showing a strong flash in SXR that resembles a 
microflare (Fig.~\ref{fig1}b). 
Meanwhile, the brightening propagated from BP1 to a remote northeast
region that is $\sim$60$\arcsec$ away, producing another bright point BP2
that is $\sim$15$\arcsec$ long and much fainter than BP1. The faint loop 
connecting BP1 and BP2 is labeled with arc1, whose length
is estimated to be 94$\arcsec$ assuming a semi-circular shape. Parallel to 
arc1, there is another arc that also takes root in BP1. Jet-like motion along 
the arc is observed. Before BP1 faded away, 
the third bright point, i.e., BP3, appeared and brightened 80$\arcsec$ east to 
BP1 (Fig.~\ref{fig1}c). It is $\sim$20$\arcsec$ in size and is more diffused
compared to BP1. Interestingly, BP3 is not an isolated one, but is probably 
connected to BP1 by arc2 in the panel. We speculate that the arc parallel
to arc1 is the right segment of arc2.
After the transient brightenings lasting for $\sim$30 min, the three 
bright points and two arcs gradually disappeared in the SXR images 
(Fig.~\ref{fig1}d).

\begin{figure}
\includegraphics[width=12cm,clip=]{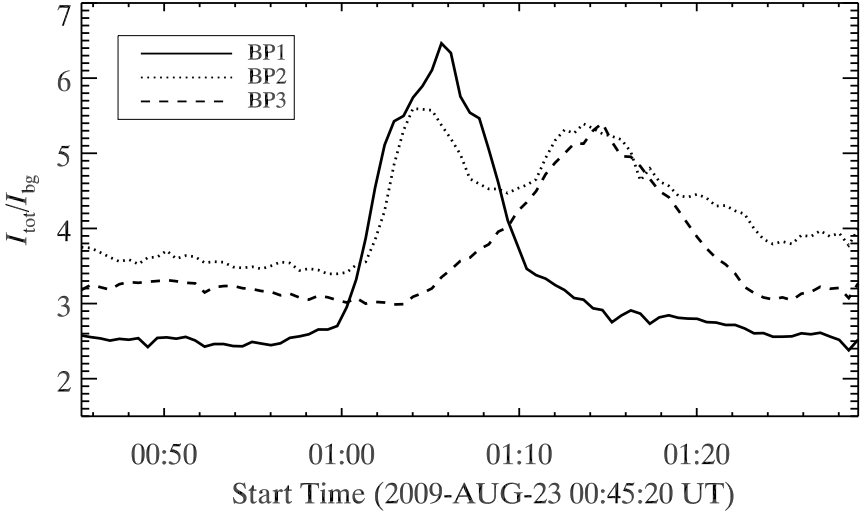}
\caption{SXR light curves of the three CBPs: 
BP1 ({\it solid line}), BP2 ({\it dotted line}), and BP3 ({\it dashed line}). 
$I_\mathrm{tot}$ and $I_\mathrm{bg}$ denote the total intensities of the 
CBPs and background region. The intensities of BP2 and BP3 are 
multiplied by 6 and 2 to get a better comparison with BP1.}
\label{fig2}
\end{figure}

Figure~\ref{fig2} displays the light curves of BP1 ({\it solid line}), 
BP2 ({\it dotted line}), and BP3 ({\it dashed line}) 
where $I_\mathrm{tot}$ and $I_\mathrm{bg}$ denote the 
total intensities of the CBPs and background region. Therefore, 
$I_\mathrm{tot}/I_\mathrm{bg}$ signifies background-normalized intensity. 
Since BP2 and BP3 are weaker than BP1, we multiply the values of 
$I_\mathrm{tot}/I_\mathrm{bg}$ by 6 and 2 for BP2 and BP3 to get a 
clearer comparison. It is revealed that the intensity of BP1 increases 
gradually from $\sim$00:56 UT to $\sim$01:00 UT before a sharp rise 
until the maximum at $\sim$01:06 UT. Then, it declines rapidly to the 
value before the onset of flash. The intensity of BP3 starts to grow at 
$\sim$01:03 UT and reaches the apex at $\sim$01:15 UT before recovering 
to a low level at $\sim$01:25 UT. The maximum of BP3 is delayed by 
$\sim$9 min compared to BP1. The intensity of BP2 begins to rise 
gently at $\sim$01:00 UT and peaks at $\sim$01:04 UT before declining to 
a lower level. The seemingly second peak at $\sim$01:14 UT is due to the 
brightening of arc2 that lies in front of BP2. Afterwards, it decays 
slowly until $\sim$01:25 UT.

\begin{figure}
\includegraphics[width=12cm,clip=]{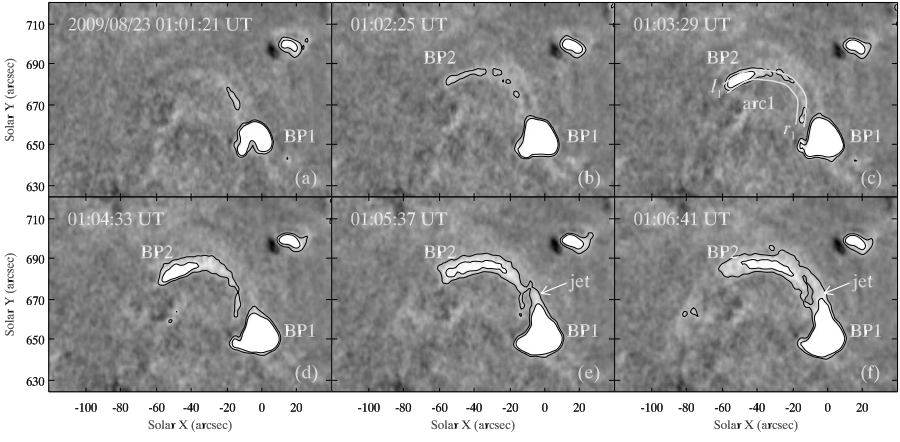}
\caption{Six snapshots of the base-difference images. The
white/black color indicates enhancement/depress in brightness.
The black contours correspond to intensities
of 25 and 50 DN s$^{-1}$.
In panel {\bf c)}, arc1 outlined by a pair of white parallel lines 
connects the two endpoints $l_1$ in BP2 and $r_1$ in BP1.
In panels {\bf e)} and {\bf f)}, the arrows point to the 
upward jet-like motion originating from BP1.}
\label{fig3}
\end{figure}

Albeit small compared to the flaring arcades (McKenzie \& Hudson 
\cite{mck99}), we discovered the signature of chromospheric evaporation
along the arcs, i.e., arc1 and arc2. To show the evaporation more 
obviously, we take the first image at 00:45:20 UT on August 23 as the 
base image before performing base-difference for the rest ones. 
Figure~\ref{fig3} displays 6 snapshots of the base-difference images 
where white/black color means intensity enhancement/depress. 
The black contours correspond to intensities of 25 and 50 DN s$^{-1}$.
In Fig.~\ref{fig3}c, a pair of white parallel curves connecting $l_1$
and $r_1$ outline arc1. It is seen from the upper panels that the
brightening propagates from $r_1$ to $l_1$ along the arc. The 
brightening does not stop abidingly at $l_1$, i.e., BP2. On the 
contrary, it propagates backwards after the brightness of BP2 reaches 
maximum at $\sim$01:04 UT, which is illustrated in the lower panels
of Fig.~\ref{fig3}.

\begin{figure}
\includegraphics[width=12cm,clip=]{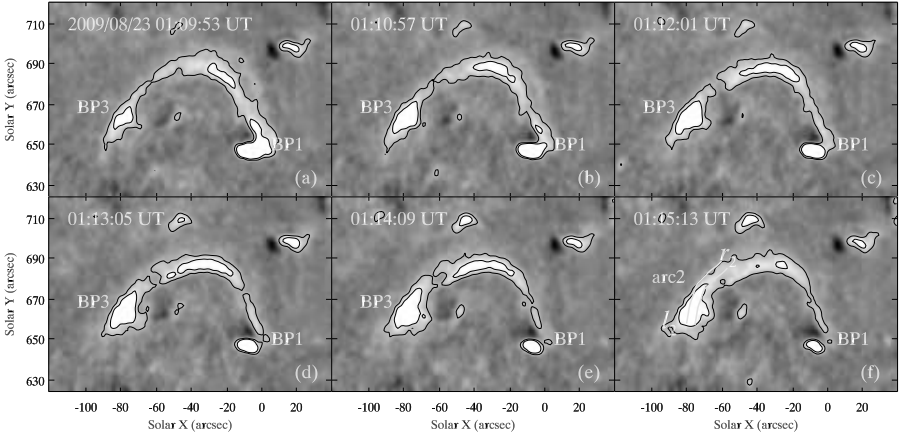}
\caption{Same as Fig.~\ref{fig3}. In panel {\bf f)}, arc2 outlined 
by a pair of white parallel lines connects the two endpoints $l_2$ 
and $r_2$.}
\label{fig4}
\end{figure}

Similar to Fig.~\ref{fig3}, we demonstrate 6 snapshots of the 
base-difference images in Fig.~\ref{fig4}. Likewise, we outline arc2
with a pair of white parallel curves connecting $l_2$ and $r_2$ in
Fig.~\ref{fig4}f. It is seen that as BP3 increases in size and 
brightness, the brightening propagates from $l_2$ to $r_2$ along arc2
during 01:10 UT -- 01:15 UT, a clear signature of chromospheric 
evaporation.

\begin{figure}
\includegraphics[width=12cm,clip=]{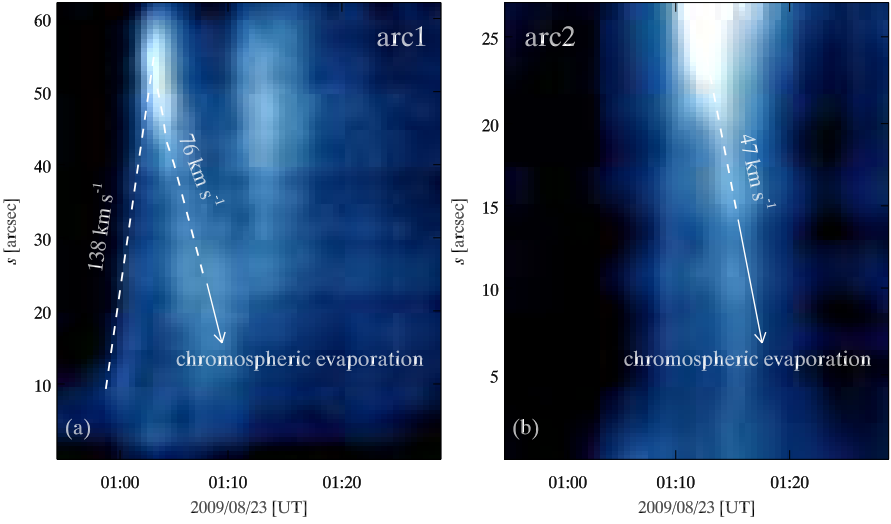}
\caption{Temporal evolutions of the SXR intensities of arc1
({\it left}) and arc2 ({\it right}) labeled in Fig.~\ref{fig3}c
and Fig.~\ref{fig4}f, respectively. The start points of 
coordinate $s$ in the $y$-axes are the right endpoints of arc1
and arc2, i.e., $r_1$ and $r_2$.
The slopes of the white dashed lines stand for the apparent 
velocities of the propagation of brightening along the arcs.
The arrows point to the features of chromospheric
evaporation.}
\label{fig5}
\end{figure}

To calculate the apparent velocities of the propagations of the 
brightenings along the arcs, we derive the temporal evolutions of SXR 
intensity of the arcs marked in Fig.~\ref{fig3}c and Fig.~\ref{fig4}f
using the standard program {\it plot\_arc.pro} in the Solar Software. 
The time-slice diagrams are presented in Fig.~\ref{fig5}, where $s$ 
denotes the distances from the right endpoints of the arcs. In the 
left panel, it seems that the brightening propagates from near the right 
endpoint during 00:58 UT -- 01:03 UT, which is outlined by the white 
dashed line. The intensity contrast between this feature and the 
background levels elsewhere is quite low. The slope of the dashed 
line provides a rough estimation of the apparent velocity of the 
propagation, i.e., 138 km s$^{-1}$. Afterwards, the brightening 
propagates reversely during 01:03 UT -- 01:07 UT. 
As we have mentioned before, it is an indication 
of chromospheric evaporation demonstrated in the lower panels of
Fig.~\ref{fig3}. The apparent velocity is estimated to be 76 km 
s$^{-1}$. In the right panel, the evaporation occurs during 
01:10 UT -- 01:15 UT at a speed of $\sim$47 km s$^{-1}$.

\section{Discussion and conclusion} \label{s-disc}

As for the mechanism of energy transport from the main flare/CBP
to sympathetic flare/CBP, several agents have been proposed in
previous works. The most popular one is nonthermal electrons 
accelerated in the main flares and guided by the large-scale
coronal loops connecting the primary and remote regions 
(Tang \& Moore \cite{tang82}; Nakajima et al. \cite{naka85}; 
Martin \& Svestka \cite{mart88}). The electrons produce chromospheric 
evaporation after penetrating into the dense chromosphere of the 
remote region. Another candidate of energy transport is thermal 
conduction at a speed of hundreds of km s$^{-1}$ due to the super-hot 
plasma created by the primary flares (Rust et al. \cite{rust85};
Bastian \& Gary \cite{bast92}). The third one is hot jets guided by 
the coronal loops as a result of interaction between small emerging 
loops and large preexisting loops 
(Hanaoka \cite{hana96,hana97}; Nishio et al. \cite{nis97}).

In our case study of sympathetic CBPs, we found tentative evidence 
of propagation of brightening from BP1 to BP2 at a speed of $\sim$138 km 
s$^{-1}$ that is far less than the typical velocity of energetic 
electrons but in the same order of magnitude as thermal conduction 
speed (Campbell \cite{cam84}; Mandrini et al. \cite{man96}). Therefore, 
we postulate that BP2 is heated by thermal conduction. For 
BP3, it is seen from Fig.~\ref{fig2} that the onset time of BP3 lags 
behind that of BP1 by $\sim$4 min, and the peak time of BP3 intensity 
is delayed by $\sim$9 min. However, definite propagation of brightening 
from BP1 to BP3 is not well observed except the jet-like motion guided 
by the arc parallel to arc1 prior to the onset of evaporation from BP3, 
as shown in Fig.~\ref{fig3}e and 
Fig.~\ref{fig3}f. If BP3 is heated by thermal conduction,
the velocity is estimated to be $\sim$380 km s$^{-1}$.

Although chromospheric evaporation have been extensively observed 
and investigated, direct evidence of the dynamic process is rare 
in SXR. Nitta (\cite{nit12}) recently reported the detection of
chromospheric evaporation upflows in transient brightening events.
The apparent velocities are up to hundreds of km s$^{-1}$. In our
case, the velocities are 76 km s$^{-1}$ and 47 km s$^{-1}$ for BP2
and BP3, respectively. Considering the velocities of 
evaporation in the sympathetic CBPs, we conclude that they belong 
to the gentle type. Milligan (\cite{mill08}) studied a 
B-class flare. There was no detectable HXR emission in the 
flare ribbons, and the blueshift was very weak ($\sim$14 km s$^{-1}$). 
Considering the apparent flow of hot material along the flare loop, 
the author tentatively proposed that it was signature of gentle
evaporation caused by thermal conduction.
We observed apparent evaporation flows of hot plasma along arc1 
and arc2 that connect BP1 with BP2 and BP3, respectively. We 
also present tentative evidence of thermal conduction front.

In this letter, we report our first discovery of sympathetic CBPs and 
the chromospheric evaporation observed by Hinode/XRT on 2009 August 
23. A strong flash occurred in the primary CBP, i.e., BP1, triggering
the appearance of two adjacent CBPs, i.e., BP2 and BP3 that are 60$\arcsec$ 
and 80$\arcsec$ away from BP1, respectively. The peak time of BP2 
intensity coincides roughly with that of BP1, but the peak time of BP3 
intensity was delayed by $\sim$9 min compared to 
BP1. The SXR brightening propagates from BP1 to 
BP2 along arc1 at a speed of $\sim$138 km s$^{-1}$ followed by 
chromospheric evaporation in BP2 at a speed of $\sim$76 km s$^{-1}$.
The propagation of brightening from BP1 to BP3 was not well identified
except the jet-like motion along the arc parallel to arc1. 
Nevertheless, we found chromospheric evaporation at a speed of 
$\sim$47 km s$^{-1}$ in BP3. We propose that the gentle evaporation
in BP2 and BP3 were generated by thermal conduction. Additional case
studies using high-cadence and high-resolution multi-wavelength 
observations are expected in the future. Numerical modelling 
would be helpful in trying to reproduce the observed upflows in response 
to a thermal conduction front.

\begin{acknowledgements}
The authors are grateful to the referee for enlightening comments
and suggestions to improve the quality of our work.
Q. M. Zhang appreciates P. F. Chen, C. Fang, M. D. Ding, H. Peter, 
J. X. Wang, W. Q. Gan, Y. P. Li, S. M. Liu, L. Feng, H. Li, and Z. J. Ning 
for valuable discussions and suggestions. Hinode is a Japanese mission 
developed and launched by ISAS/JAXA, with NAOJ as domestic partner 
and NASA and STFC (UK) as international partners. The research is 
supported by the Chinese foundations NSFC (11025314, 10878002,  
10933003, and 11173062) and 2011CB811402.
\end{acknowledgements}

\end{document}